\documentclass[12pt, letterpaper]{article}

\usepackage{amsmath,amssymb}
\usepackage{cite}
\usepackage{fancyhdr}
\usepackage[top=1in, bottom=1in, left=1in, right=1in]{geometry}
\usepackage{graphicx}
\usepackage{hyperref}

\numberwithin{equation}{section}
\date{}

\title{{\rm\footnotesize \qquad \qquad \qquad \qquad \qquad \ \qquad \qquad \qquad \ \ \ \ \ \                      RUNHETC-2023-27}\vskip.5in     
My Personal History With the Quantum Theory of de Sitter Space    }
\author{Tom Banks\\
NHETC and Department of Physics \\
Rutgers University, Piscataway, NJ 08854-8019\\
E-mail: \href{mailto:tibanks@ucsc.edu}{tibanks@ucsc.edu}
\\
\\}
\begin{document}

\maketitle
\thispagestyle{fancy} 

\begin{abstract}  The purpose of this brief article is to clarify certain distortions of the history of ideas about the quantum theory of de Sitter space that have appeared in recent literature.

\normalsize \noindent  \end{abstract}


\newpage
\tableofcontents
\vspace{1cm}

\vfill\eject
\section{Introduction}

I first heard about the cosmological constant problem from K. Johnson in a lecture at the Scottish Summer School in St. Andrews in 1976.  L. Susskind, who had emphasized the importance of the gauge hierarchy problem to me since we first worked together in 1975, often referred to it as an even bigger mystery.  In 1999 I was sitting on the patio of my new home in Big Sur, staring at the ocean, when a number of thoughts came together in my head.   We had learned from AdS/CFT that when it was negative, the c.c. should be thought of as a parameter characterizing the high energy behavior of particular models, and was not renormalized in low energy effective field theory.  My friends Fischler and Susskind had proposed generalizing the Bekenstein-Hawking bound for black holes to cosmological space-times\cite{fs}, and this had led Bousso to a general conjecture on entropy bounds for space-time regions\cite{bousso}.  The universe we inhabited seemed to be approaching a de Sitter asymptotic future, and Gibbons and Hawking had predicted that such a universe had finite entropy.   Moreover, conventional string theory seemed unable to cope with such a universe.

Suddenly it all made sense.  The Gibbons-Hawking entropy should be thought of as the log of the dimension of the Hilbert space, which implied that empty dS space was a maximal entropy state, and indeed from the point of view of any geodesic, anything not bound to the geodesic appears to fade into its cosmological horizon in a time of order the dS radius.   Furthermore, the black hole entropy formula showed that objects bound to the geodesic {\it lowered} the entropy\footnote{It wasn't until somewhat later that I realized that this gave a derivation of the precise value of the Gibbons-Hawking temperature independent of quantum field theory (QFT).   It was much later that Fischler and I realized that this was the key clue to how local physics arises from a quantum theory of horizon degrees of freedom.}.   

I spoke about these ideas at the 60th birthday celebration for L. Susskind at Stanford in the Spring of 2000.  Fischler was there and told me he'd had the same idea independently.  He spoke about his version at Geoff West's 60th birthday celebration in Taos, NM, also in 2000.  I've always tried to give him equal credit, but the community has resisted this.  I finally published a version of this in connection with the talk I gave at the strings conference in Michigan\cite{tb2001}.   Some people\cite{bousso2001} greeted the idea that dS space had a finite dimensional Hilbert space with a certain amount of enthusiasm, but for the most part it was met with skepticism and mockery.  I spent hours trying to convince E. Witten that the idea was correct, despite the fact that in global coordinates the past and future asymptotic regions of dS space seemed to have arbitrarily large space-like slices.   

I gave two arguments, one mathematical, one physical.  The mathematical argument was that because there were no asymptotic space-like or null regions, de Sitter isometries were all gauge transformations and all of dS space consisted of gauge copies of a single static patch.   The physical argument was that generic asymptotic data from the past or future extrapolated to a singularity that did not allow one to continue to the future or past of dS, so that specifying both past and future dS asymptotics restricted one to a finite dimensional Hilbert space.  Gary Horowitz told me this sounded plausible, and Bousso and Freivogel\cite{bfrei} eventually proved a theorem along these lines.   Witten seemed to partially concede in\cite{witten2001}.

The conjecture that dS space has a finite number of states gets reinforced by the computation of Coleman DeLuccia tunneling amplitudes between two dS spaces\cite{cdl}, which obey a rule of detailed balance\cite{heretics} appropriate to tunneling in finite systems.  This leads to a controversial reinterpretation of the theory of eternal inflation\cite{EI}, with by far the most sensible conjecture about the meaning of transitions to Big Crunch cosmologies.  It turns the search for a measure on an ill defined infinite probability space into an exercise in the non-equilibrium statistical mechanics of finite quantum systems.  

In the meantime, Fischler and I continued to develop these ideas into a general formalism for dealing with cosmological space-times quantum mechanically.  Eventually we gave it the name Holographic Space-time.  The central idea was to generalize the conjecture about dS space to a conjecture about the density matrix of every finite area causal diamond in "empty space".   Bousso pointed out to us that he had made a similar conjecture in his review papers\cite{boussoreview}.  This led us to a model of the maximally entropic universe, which is a flat FRW universe with equation of state $p = \rho$\cite{mannelli}.  If we had been more clear thinking, we would have realized that this model in fact {\it contradicted} our initial ansatz for the density matrix.  In fact, to fit the data of the $p = \rho$ universe, we had to assume that the modular Hamiltonian at any time was that of a $1 + 1$ dimensional conformal field theory (CFT), with central charge that grew like the area of the apparent horizon.  We did notice the connection to earlier work of Carlip\cite{Carlip} on black hole horizons, but did not pay enough attention.  And it was only much later that we realized that simply stopping the expansion of the Hilbert space of our model with a fixed CFT, gave a description of the flat FRW model with
$ a(t) = \sinh^{1/3} (3t/R)$, which asymptotes to dS space.   This implies that the density matrix of dS space is {\it not} proportional to the unit matrix.   It does not affect the ability to interpret the dS temperature in terms of a connection between energy measured along a geodesic, and the number of constraints on boundary degrees of freedom\cite{tbpd}.  

Fischler and I have always been loath to use the term observer because of all of the nonsensical literature about the connection between quantum measurement theory and consciousness.  Instead, the HST formalism has always been based on the Hilbert spaces associated with intervals along time-like trajectories.  One thinks of these as the trajectories of detectors that have no back-reaction on space-time geometry.  I'll discuss the problem of back-reaction below. The covariant entropy bound implies that for short enough proper time the maximal entropy causally accessible to a detector on such a trajectory is always finite, which means expectation values of operators are computable in a Type $II_1$ von Neumann algebra.  The hyperfinite Type $II_1$ algebra is just the limit of a finite number of fermionic creation and annihilation operators.  For a black hole of mass $1$ gram in $4$ dimensions, the entropy is approximately $10^{11}$, so one must be able to do measurements fine enough to distinguish between the infinite dimensional algebra and that of $10^{11}$ q-bits in order to falsify the claim of HST that the Hilbert space is finite dimensional.  For the universe as a whole the number of q-bits is about $10^{123}$ and it's clearly impossible to imagine making the distinction. As we'll see, realistic measurements inside a causal diamond cannot actually probe even a finite fraction (parametrically) of the quantum information in the diamond.   Thus, we've always assumed finite dimensional Hilbert spaces, which also implies a discreteness of time evolution. We take the discrete interval to be the Planck time.  

A key feature of the formalism has always been the Quantum Principle of Relativity (QPR) (which should have been a clue to the tone deaf that we were talking about "observers").  Any given space-time is described by multiple quantum systems, living in a Hilbert bundle over a manifold of time-like trajectories.  The simplest collection of time-like trajectories that form a manifold is the space of time-like geodesics.  In examples like AdS space one might want to add null-geodesics or accelerated trajectories that reach the boundary.   For any two choices of interval along a pair of trajectories in the base space of the bundle there is an overlap region and a causal diamond, unique up to symmetries, of maximal area in that region.  We insist that there be a Hilbert space associated with this diamond and unitary embedding of this space into the Hilbert spaces of each individual trajectory as a tensor factor.  The density matrices on those two tensor factors must have identical entanglement spectra.  This contrains both the time dependent Hamiltonians and the choices of initial state on all fibers of the Hilbert bundle.

This formulation of HST depends on the assumption of a background space-time.  The proper way to think about this is using Jacobson's\cite{ted95} demonstration that Einstein's equations (doubly projected onto an arbitrary null vector so that the c.c. doesn't contribute) are the hydrodynamic equations of the area law for causal diamonds.  Thus, we view a space-time as a hydrodynamic background, into which we must fit a quantum theory whose hydrodynamics matches that background.  The refinement of Jacobson's argument due to Carlip\cite{Carlip} and Solodukhin\cite{solo}, generalized from black holes to arbitrary diamonds in\cite{BZ}, shows us how to do this.  These authors show that the fluctuations of the conformal factor of the background geometry of the holographic screen of each diamond are the hydrodynamic equations of a $1 + 1$ dimensional CFT with central charge proportional to the area of the diamond's holographic screen.  They derive the area law for the diamond entropy, assuming the modular Hamiltonian is the $L_0$ generator of that CFT.   The HST formalism provides us with a guess for what the CFT is\cite{hilbertbundles}.  

\section{Locality}

Perhaps the most important insight into the relation between the quantum theory of gravity and quantum field theory comes from Fiol's explicit modeling of the entropy reduction of localized objects in dS space\cite{bfm}.  The variables are rectangular matrices, each matrix element being a $1 + 1$ dimensional massless fermion field (with a UV cutoff described in\cite{hilbertbundles}) .   Localized states are states on which matrices bilinear in the fermions are block diagonal.   This prescription reproduces the thermal properties of dS space\cite{bfm}\cite{tbpd}, but also leads to a model of scattering of particles in Minkowski space in the limit of infinite dS radius\cite{tbwfscatt}. 
This is an alternative view of the emergence of local physics from a holographic theory of quantum gravity, which shows how locality breaks down at the threshold of black hole formation.  It has puzzled me that so little attention has been paid to it.   By using tensor network technology, it can be made compatible with the standard tensor network picture of locality in AdS space\cite{tbwfads}, on scales larger than the AdS radius.  Although the original implementation of Fiol's idea was based on the assumption that the density matrix of empty dS space was proportional to the unit matrix, it works just as well, with one proviso, for the Carlip-Solodukhin ansatz\cite{tbpd}.  The proviso is that we must give up the requirement that the probability distribution for states that have energy (as measured along a geodesic in the diamond) of order $1/R_{\diamond}$ is just given by the ratio of the dimension of the constrained subspace in which they live, to the total.  This remains true for those low energy states that can actually be detected on the geodesic, and are well described by QFT, but is not true for most of the states on the boundary.   

\subsection{The Limitations of Quantum Field Theory}

There is a widespread folklore in the string theory community that in various "weak string coupling" or "large AdS radius" limits, QFT is a good approximation to everything in a model of quantum gravity.   This is simply incorrect and has been known to be incorrect since at least the 1970s.   If one considers a causal diamond in Minkowski space of proper time $T$, the field theory Hilbert space one associates with it has a dimension which is the exponential of $(TM_c)^{d-1}$, where $M_c$ is some kind of UV cut-off.  Most of these states have semi-classical gravitational back-reaction that produces black holes with size larger than $T$.  Eliminating those, one is left, generically, with an entropy bounded by something of order $(A_{\diamond}/G_N)^{\frac{d-1}{d}}$.   In particular, one has to omit the area's worth of states that give rise to the UV divergent state independent QFT entanglement entropy of the diamond\footnote{Note that in a pre-existing black hole background, the near horizon states have smaller gravitational back reaction due to the gravitational redshift.  This is the reason they've been viewed as candidates for the explanation of black hole entropy.  Making this assumption leads to the firewall paradox.}.  In an important paper\cite{CKN}, Cohen, Kaplan and Nelson, showed that omitting these states did not affect the agreement between QFT and any known experiment.  The intuitive reason for this is that experiments are done on near geodesic trajectories for finite proper times, and are relatively unaffected by virtual excitations (in QFT language) far from those trajectories.  CKN used a crude correlation between UV and IR cutoffs and found that their bounds were close to current experimental limits.  More sophisticated cutoff procedures\cite{draperetal} allow more high momentum modes near the experimental trajectory and put experimental probes of the bound further out of reach.   

The upshot of these arguments is that we have no phenomenological reason to believe the QFT picture of most of the states on the boundaries of a causal diamond.  Since that picture leads to the firewall problem, and does not account microscopically for the large increase in entropy when a low entropy object is dropped onto a large black hole (whereas the HST models referred to above avoid the firewall and do explain the entropy increase), one should simply accept the fact that QFT breaks down, not just at large energy and short distance, but in any situation with large entropy.  There are indications from perturbative string theory\cite{raju} that this is the case.

\section{The Back Reaction of Detectors}

According to the modern rules of quantum measurement theory, a detector that has to measure and store the quantum information about a quantum system with $M$ dimensions has to have $M$ independent "collective variables" with the following properties
\begin{itemize}
\item There is a large integer $N$ such that the uncertainties of all of the collective variables scale like $N^{-1}$.  The equations of motion of the detector system relate collective variables to each other, so we can follow histories of these variables with no ambiguity, for times of order $N$.
\item The interference terms in the probability sum rule for different histories of the collective variables are exponentially small, of order $ e^{- N^p}$ with $p$ of order $1$ or larger. 
\item The detector's collective variables must retain their "semi-classical" status over time scales $N$ long enough to do multiple measurements on different bases of the $M$ dimensional system, in order to be able to extract all of the probability amplitudes for different time dependent transitions.  
\end{itemize}
It follows from this that the detector must be a quantum system with a Hilbert space whose dimension is $\gg M$.  Just the requirement that the detector be a subsystem of a closed finite entropy universe makes it impossible to imagine a detector inside a closed universe that can verify every detail of the quantum model of the universe.  However, the features of the quantum theory of gravity discussed above make the problem much more severe than this.

First, all experiments to date have taken place on near geodesic trajectories in a causal diamond.  The more complex a detector we imagine on such a trajectory, the more constrained the state of the boundary quantum system it is measuring.  Thus, not only can't we imagine a detector with enough collective variables to determine the quantum behavior of the universe, but the more precise and efficient our detector, the less of the Hilbert space of the universe the detector can explore.

Furthermore, the kinds of systems that we know to exist on near geodesic trajectories are those described by QFT, and black holes.   We've argued that, if our universe is a dS space with the observed value of the c.c. then the log of the dimension of the Hilbert space of the maximal QFT detector allowed by theory\footnote{Note that the actual entropy of localized matter in our universe is about $10^{90}$ and is dominated by the entropy of supermassive black holes in the centers of galaxies.} is about $10^{92}$, while that of the universe is $10^{123}$.  Black holes have more quantum states for a given entropy deficit on the horizon, but they have many fewer collective coordinates with which to record and store data.   So the quantum gravity constraints on detectors inside a dS universe imply that much of the quantum information in the universe is undetectable. 

The combination of quantum measurement theory and the constraints of quantum gravity thus imply that no observations in a dS universe can ever test most of the details of a quantum model of the entire universe.  Note that we have not yet taken into account the constraint that the detector's collective coordinates survive long enough to make all the required measurements.   If we imagine a detector moving along a geodesic, these constraints are very severe.   Most time-like trajectories approach the boundary of the causal diamond of any given geodesic in a time of order the dS radius.   Quantum fluctuations of the detector's position, and disturbances of its position by gravitational scattering from other localized matter, Gibbons-Hawking radiation {\it etc.} will make it deviate from geodesic motion on time scales short compared to those necessary to the collection of the maximal amount of quantum information it can process.  Fortunately, for detectors that belong to local groups of galaxies like our own, geodesic motion is irrelevant.  The huge set of quantum states of the local group decohere the trajectory of its center of mass motion through the universe.   In principle then, the useful lifetime of such a massive complex detector is limited by the time it takes it to collapse into a black hole.   After that point, the information it has collected has "unhappened", it is lost in the detailed quantum microstate of the black hole, and no macroscopic record remains.

\section{Conclusions}
I list the main take-aways from this article
\begin{itemize}
\item The original claim that the density matrix of empty dS space is maximally uncertain was made independently by myself and W. Fischler in 2000.  The most important part of that claim was the identification of the energy of localized states as a count of the number of constrained q-bits in the subspace of fixed localized energy.  Fiol invented a fermionic matrix model of this effect\cite{bfm}, and it leads to an understanding of how local particle physics emerges from a holographic model of gravity in small causal diamonds.  This is quite different from the way that locality in the AdS part of space-time emerges in the tensor network approach to the AdS/CFT correspondence.
\item Operator algebras/Hilbert spaces of finite area diamonds are finite dimensional, and time evolution is consequently discrete, consisting of entangling of larger and larger tensor factors of a large Hilbert space with the rest of the degrees of freedom.  The quantum of time is taken to be the Planck scale.   
\item The maximal uncertainty ansatz is probably wrong and should be replaced by the Carlip-Solodukhin ansatz that the modular Hamiltonian of a diamond is the (appropriately cut-off) Virasoro generator $L_0$ of a $1 + 1$ dimensional CFT living on the stretched horizon of the diamond.   This is valid for all causal diamonds in space-times with non-negative c.c..   For negative c.c. the C-S ansatz is valid for boundary anchored Ryu-Takayanagi diamonds, and for diamonds smaller than the AdS radius.   Diamonds larger than the AdS radius must be treated by the tensor network construction of\cite{tbwfads} and have the modular Hamiltonian of a thermal CFT in higher dimensions.  For small causal diamonds or diamonds in space-times with non-negative c.c., the connection between local physics and boundary constraints, and the derivation of the dS temperature from this relation, are compatible with the C-S ansatz as long as we restrict our assumptions about the range of validity of QFT in the bulk, in the manner suggested by CKN.  A particularly welcome consequence of abandoning the maximal uncertainty hypotheses is that it implies fluctuations of the modular Hamiltonian in inflationary horizon volumes.  These give rise to the observed temperature fluctuations in the Cosmic Microwave Background. A model without them would be ruled out by observation.
\item QFT is valid in a causal diamond at most for states of entropy bounded by $(A/G_N)^{\frac{d-1}{d}} $.
\item The "vacuum" in Minkowski space is a high entropy state, with the entropy coming from zero energy states.  This follows by viewing Minkowski space as a limit of dS space, but also as a limit from negative c.c..  In the AdS case, the Minkowski diamond is a tiny subsystem entangled with the much larger AdS background, as seen in the tensor network formulation.  In correlation functions the entanglement comes from Witten diagrams in which arbitrary numbers of massless lines are exchanged between operators focussed on the Minkowski diamond (which converge to Minkowski S-matrix elements with finite numbers of particles) and operators that create excitations that propagate far from the diamond.  The nature of the asymptotic Hilbert space and the existence of a non-perturbative S-matrix remain open, particularly in $4$ dimensions.
\item The joint constraints of quantum measurement theory and quantum gravity imply that no detector in a closed universe can probe most of the quantum information that the covariant entropy bound allows in that universe.  Consequently, no model of such a universe with infinite dimensional algebras of observables has any observational meaning, and even finite dimensional models will have a lot of freedom that cannot be verified or falsified by experiment.  
\end{itemize}

For those with endless patience, a fairly complete list of my thoughts on the entropy of states in dS space and their wider implications, full of mistakes and false starts, can be found here\cite{holocosm}.  
\vskip.3in
\begin{center}
{\bf Acknowledgments }
\end{center}
 The work of T.B. was supported by the Department of Energy under grant DE-SC0010008. Rutgers Project 833012.  I would like to thank W. Fischler for collaboration over many years.  He deserves equal credit for most of the insights about entropy, space-time and locality that we've explore in the past 25 years.  The opinions expressed in this paper are my own.



\begin{thebibliography}{99}
   \bibitem{tb2000}T.~Banks, Talk at the 60th Birthday Celebration of L. Susskind, Stanford University, 2000 ; T.~Banks,
``Cosmological breaking of supersymmetry?,''
Int. J. Mod. Phys. A \textbf{16}, 910-921 (2001)
doi:10.1142/S0217751X01003998
[arXiv:hep-th/0007146 [hep-th]].
\bibitem{wf2000} W. Fischler, {\it Taking de Sitter Seriously}, talk at the 60th Birthday Celebration for G. West, Taos, NM, 2000.
\bibitem{bfm}T.~Banks, B.~Fiol and A.~Morisse,
``Towards a quantum theory of de Sitter space,''
JHEP \textbf{12}, 004 (2006)
doi:10.1088/1126-6708/2006/12/004
[arXiv:hep-th/0609062 [hep-th]].


  
   \bibitem{fs}  W.~Fischler and L.~Susskind,
  ``Holography and cosmology,''
  hep-th/9806039;
 \bibitem{bousso} R.~Bousso,
  ``A Covariant entropy conjecture,''
  JHEP {\bf 9907}, 004 (1999)
  [hep-th/9905177].
\bibitem{tb2001} T.~Banks, ``Cosmological breaking of supersymmetry?,''
Int. J. Mod. Phys. A \textbf{16}, 910-921 (2001)
doi:10.1142/S0217751X01003998
[arXiv:hep-th/0007146 [hep-th]].
\bibitem{bousso2001} R.~Bousso, ``Positive vacuum energy and the N bound,''
JHEP \textbf{11}, 038 (2000)
doi:10.1088/1126-6708/2000/11/038
[arXiv:hep-th/0010252 [hep-th]].
\bibitem{witten2001} E.~Witten, ``Quantum gravity in de Sitter space,''
[arXiv:hep-th/0106109 [hep-th]].
\bibitem{cdl} S.~R.~Coleman and F.~De Luccia,``Gravitational Effects on and of Vacuum Decay,''
Phys. Rev. D \textbf{21}, 3305 (1980)
doi:10.1103/PhysRevD.21.3305
\bibitem{heretics} T.~Banks,``Heretics of the false vacuum: Gravitational effects on and of vacuum decay. 2.,''
[arXiv:hep-th/0211160 [hep-th]];
A.~R.~Brown and E.~J.~Weinberg, ``Thermal derivation of the Coleman-De Luccia tunneling prescription,''
Phys. Rev. D \textbf{76}, 064003 (2007)
doi:10.1103/PhysRevD.76.064003
[arXiv:0706.1573 [hep-th]].
\bibitem{EI} T.~Banks,``Relaxation of the Cosmological Constant,''
Phys. Rev. Lett. \textbf{52}, 1461-1463 (1984)
doi:10.1103/PhysRevLett.52.1461;
T.~Banks and M.~Johnson,``Regulating eternal inflation,''
[arXiv:hep-th/0512141 [hep-th]];
A.~Aguirre, T.~Banks and M.~Johnson,``Regulating eternal inflation. II. The Great divide,''
JHEP \textbf{08}, 065 (2006)
doi:10.1088/1126-6708/2006/08/065
[arXiv:hep-th/0603107 [hep-th]];
T.~Banks and J.~F.~Fortin,``Tunneling Constraints on Effective Theories of Stable de Sitter Space,''
Phys. Rev. D \textbf{80}, 075002 (2009)
doi:10.1103/PhysRevD.80.075002
[arXiv:0906.3714 [hep-th]];
T.~Banks,``On the Limits of Effective Quantum Field Theory: Eternal Inflation, Landscapes, and Other Mythical Beasts,''
[arXiv:1910.12817 [hep-th]];
T.~Banks and B.~Zhang,``Comment on Coleman-DeLuccia instantons,''
SciPost Phys. \textbf{12}, no.1, 015 (2022)
doi:10.21468/SciPostPhys.12.1.015
[arXiv:2106.12696 [hep-th]].
\bibitem{bfrei}R.~Bousso and B.~Freivogel, ``Asymptotic states of the bounce geometry,''
Phys. Rev. D \textbf{73}, 083507 (2006)
doi:10.1103/PhysRevD.73.083507
[arXiv:hep-th/0511084 [hep-th]].
\bibitem{boussoreview} R.~Bousso, ``Holography in general space-times,''
JHEP \textbf{06}, 028 (1999)
doi:10.1088/1126-6708/1999/06/028
[arXiv:hep-th/9906022 [hep-th]];
R.~Bousso, ``The Holographic principle for general backgrounds,''
Class. Quant. Grav. \textbf{17}, 997-1005 (2000)
doi:10.1088/0264-9381/17/5/309
[arXiv:hep-th/9911002 [hep-th]].
 \bibitem{ted95}T.~Jacobson,
``Thermodynamics of space-time: The Einstein equation of state,''
Phys. Rev. Lett. \textbf{75}, 1260-1263 (1995)
doi:10.1103/PhysRevLett.75.1260
[arXiv:gr-qc/9504004 [gr-qc]].
\bibitem{mannelli} T.~Banks, W.~Fischler and L.~Mannelli,
  ``Microscopic quantum mechanics of the p = rho universe,''
  Phys.\ Rev.\ D {\bf 71}, 123514 (2005)
  [hep-th/0408076];
 \bibitem{Carlip} S.~Carlip,
``Black hole entropy from conformal field theory in any dimension,''
Phys. Rev. Lett. \textbf{82}, 2828-2831 (1999)
doi:10.1103/PhysRevLett.82.2828
[arXiv:hep-th/9812013 [hep-th]].
\bibitem{solo} S.~N.~Solodukhin,
``Conformal description of horizon's states,''
Phys. Lett. B \textbf{454}, 213-222 (1999)
doi:10.1016/S0370-2693(99)00398-6
[arXiv:hep-th/9812056 [hep-th]].
\bibitem{BZ} T.~Banks and K.~M.~Zurek,
``Conformal description of near-horizon vacuum states,''
Phys. Rev. D \textbf{104}, no.12, 126026 (2021)
doi:10.1103/PhysRevD.104.126026
[arXiv:2108.04806 [hep-th]].
\bibitem{tbpd} T.~Banks and P.~Draper,
``Comments on the entanglement spectrum of de Sitter space,''
JHEP \textbf{01}, 135 (2023)
doi:10.1007/JHEP01(2023)135
[arXiv:2209.08991 [hep-th]].
\bibitem{CKN} A.~G.~Cohen, D.~B.~Kaplan and A.~E.~Nelson,
  ``Effective field theory, black holes, and the cosmological constant,''
  Phys.\ Rev.\ Lett.\  {\bf 82}, 4971 (1999)
  doi:10.1103/PhysRevLett.82.4971
  [hep-th/9803132].
\bibitem{tbwfscatt} T.~Banks and W.~Fischler, ``Holographic Theory of Accelerated Observers, the S-matrix, and the Emergence of Effective Field Theory,''
[arXiv:1301.5924 [hep-th]].
\bibitem{hilbertbundles} T.~Banks, ``Hilbert Bundles and Holographic Space-time Models,''
[arXiv:2306.07038 [hep-th]].
\bibitem{raju} S.~Ghosh and S.~Raju, ``Breakdown of String Perturbation Theory for Many External Particles,''
Phys. Rev. Lett. \textbf{118}, no.13, 131602 (2017)
doi:10.1103/PhysRevLett.118.131602
[arXiv:1611.08003 [hep-th]];
S.~Ghosh and S.~Raju,``Loss of locality in gravitational correlators with a large number of insertions,''
Phys. Rev. D \textbf{96}, no.6, 066033 (2017)
doi:10.1103/PhysRevD.96.066033
[arXiv:1706.07424 [hep-th]].

\bibitem{draperetal} T.~Banks and P.~Draper,
``Remarks on the Cohen-Kaplan-Nelson bound,''
Phys. Rev. D \textbf{101}, no.12, 126010 (2020)
doi:10.1103/PhysRevD.101.126010
[arXiv:1911.05778 [hep-th]]; N.~Blinov and P.~Draper,
``Densities of states and the Cohen-Kaplan-Nelson bound,''
Phys. Rev. D \textbf{104}, no.7, 076024 (2021)
doi:10.1103/PhysRevD.104.076024
[arXiv:2107.03530 [hep-ph]].

  \bibitem{tbwfads} T.~Banks and W.~Fischler,
``Holographic Space-time Models of Anti-deSitter Space-times,''
[arXiv:1607.03510 [hep-th]].
\bibitem{gh}G.~W.~Gibbons and S.~W.~Hawking,
``Cosmological Event Horizons, Thermodynamics, and Particle Creation,''
Phys. Rev. D \textbf{15}, 2738-2751 (1977)
doi:10.1103/PhysRevD.15.2738
\bibitem{tbpd}T.~Banks and P.~Draper,
``Comments on the entanglement spectrum of de Sitter space,''
JHEP \textbf{01}, 135 (2023)
doi:10.1007/JHEP01(2023)135
[arXiv:2209.08991 [hep-th]].

\bibitem{holocosm} T.~Banks and W.~Fischler, ``M theory observables for cosmological space-times,''
[arXiv:hep-th/0102077 [hep-th]].

 T.~Banks and W.~Fischler,
  ``An Holographic cosmology,''
  hep-th/0111142;
 T.~Banks, W.~Fischler and S.~Paban,
  ``Recurrent nightmares? Measurement theory in de Sitter space,''
  JHEP {\bf 0212}, 062 (2002)
  [hep-th/0210160];
  T.~Banks,``Some thoughts on the quantum theory of de sitter space,''
[arXiv:astro-ph/0305037 [astro-ph]];
 T.~Banks and W.~Fischler,
  ``Holographic cosmology 3.0,''
  Phys.\ Scripta T {\bf 117}, 56 (2005)
  [hep-th/0310288];
  T.~Banks,``Supersymmetry, the cosmological constant, and a theory of quantum gravity in our universe,''
Gen. Rel. Grav. \textbf{35}, 2075-2078 (2003)
doi:10.1023/A:1027342603060
[arXiv:hep-th/0305206 [hep-th]];
 T.~Banks, W.~Fischler and L.~Mannelli,
  ``Microscopic quantum mechanics of the p = rho universe,''
  Phys.\ Rev.\ D {\bf 71}, 123514 (2005)
  [hep-th/0408076];
  T.~Banks, ``Some thoughts on the quantum theory of stable de Sitter space,''
[arXiv:hep-th/0503066 [hep-th]].
T.~Banks and W.~Fischler,
  ``Holographic Theories of Inflation and Fluctuations,''
  arXiv:1111.4948 [hep-th];
 T.~Banks,
  ``Holographic Space-Time: The Takeaway,''
  arXiv:1109.2435 [hep-th];
T.~Banks and J.~Kehayias,
  ``Fuzzy Geometry via the Spinor Bundle, with Applications to Holographic Space-time and Matrix Theory,''
  Phys.\ Rev.\ D {\bf 84}, 086008 (2011)
  [arXiv:1106.1179 [hep-th]];
 T.~Banks,
  ``TASI Lectures on Holographic Space-Time, SUSY and Gravitational Effective Field Theory,''
  arXiv:1007.4001 [hep-th];
  T.~Banks, ``Entropy and initial conditions in cosmology,''
[arXiv:hep-th/0701146 [hep-th]];
T.~Banks, ``Holographic Space-time from the Big Bang to the de Sitter era,''
J. Phys. A \textbf{42}, 304002 (2009)
doi:10.1088/1751-8113/42/30/304002
[arXiv:0809.3951 [hep-th]].
 T.~Banks,
  ``Holographic space-time and its phenomenological implications,''
  Int.\ J.\ Mod.\ Phys.\ A {\bf 25}, 4875 (2010)
  [arXiv:1004.2736 [hep-th]];
 T.~Banks,
  ``Holographic Space-time from the Big Bang to the de Sitter era,''
  J.\ Phys.\ A A {\bf 42}, 304002 (2009)
  [arXiv:0809.3951 [hep-th]].
   T.~Banks and W.~Fischler,
  ``CP Violation and Baryogenesis in the Presence of Black Holes,''
  arXiv:1505.00472 [hep-th].
   T.~Banks and W.~Fischler,
  ``Holographic Inflation Revised,''
  arXiv:1501.01686 [hep-th].
T.~Banks and W.~Fischler,
``Holographic Space-time and Newton's Law,''
[arXiv:1310.6052 [hep-th]].
 \end{thebibliography}


\end{document}